

\input harvmac


\overfullrule=0pt

\def\A{{\scriptscriptstyle A}}

\def\C{{\scriptscriptstyle C}}

\def\I{{\scriptscriptstyle I}}

\def\L{{\scriptscriptstyle L}}

\def\P{{\scriptscriptstyle P}}

\def\R{{\scriptscriptstyle R}}

\def\T{{\scriptscriptstyle T}}

\def\W{{\scriptscriptstyle W}}

\def\Y{{\scriptscriptstyle Y}}
\def\Z{{\scriptscriptstyle Z}}

\def\CL{{\C\L}}
\def\LR{{\L\R}}
\def\RC{{\R\C}}


\def\CF{{\cal F}}
\def\CG{{\cal G}}

\def\CM{{\cal M}}


\def\a{\alpha}
\def\b{\beta}

\def\th{\theta}
\def\u{\mu}
\def\v{\nu}


\def\aEM{\alpha_{\scriptscriptstyle EM}}
\def\aS{\alpha_s}
\def\aten{\alpha_{\scriptscriptstyle 10}}
\def\bar#1{\overline{#1}}

\def\bone{b_{\scriptscriptstyle 1}}
\def\btwo{b_{\scriptscriptstyle 2}}
\def\btwop{b'_{\scriptscriptstyle 2}}
\def\btwoL{b_{\scriptscriptstyle 2L}}
\def\bthree{b_{\scriptscriptstyle 3}}
\def\bfour{b_{\scriptscriptstyle 4}}
\def\bfourL{b_{\scriptscriptstyle 4L}}

\def\cossq{\cos^2 \th_\W}
\def\dash{{\> \over \>}} 		

\def\EM{{\scriptscriptstyle EM}}

\def\gone{g_{\scriptscriptstyle 1}}
\def\gtwoL{g_{\scriptscriptstyle 2L}}
\def\gtwoR{g_{\scriptscriptstyle 2R}}
\def\gtwop{g'_{\scriptscriptstyle 2}}
\def\gtwoLp{g'_{\scriptscriptstyle 2L}}
\def\gtwoRp{g'_{\scriptscriptstyle 2R}}
\def\gthree{g_{\scriptscriptstyle 3}}
\def\gfour{g_{\scriptscriptstyle 4}}
\def\gfourL{g_{\scriptscriptstyle 4L}}
\def\gfourR{g_{\scriptscriptstyle 4R}}
\def\gtenL{g_{\scriptscriptstyle 10L}}
\def\gtenR{g_{\scriptscriptstyle 10R}}
\def\GeV{\>\, \rm GeV}

\def\gtap{\raise.3ex\hbox{$>$\kern-.75em\lower1ex\hbox{$\sim$}}}
\def\hc{\rm h.c.}
\def\LC{\Lambda_\C}
\def\LLR{\Lambda_\LR}

\def\ltap{\raise.3ex\hbox{$<$\kern-.75em\lower1ex\hbox{$\sim$}}}

\def\Mgut{M_{\scriptscriptstyle GUT}}
\def\mxpt{(-\vec{x},t)}
\def\Mz{M_\Z}
\def\NF{N_{\CF}}
\def\PhiL{\Phi_\L}
\def\PhiR{\Phi_\R}
\def\phiL{\phi_\L}
\def\phiR{\phi_\R}
\def\PsiL{\Psi_\L}
\def\PsiR{\Psi_\R}
\def\psiL{\psi_\L}
\def\psiR{\psi_\R}

\def\QCD{{\scriptscriptstyle QCD}}
\def\QFD{{\scriptscriptstyle QFD}}
\def\sinsq{\sin^2 \th_\W}

\def\sp{\>\,}

\def\TeV{\>\, \rm TeV}
\def\third{{1 \over 3}}
\def\Tr{\rm{Tr} \,}
\def\vL{v_\L}
\def\vR{v_\R}
\def\xpt{(\vec{x},t)}
\def\yrs{\>\, \rm{yrs}}



\def\fourthirds{{4 \over 3}}
\def\half{{1 \over 2}}

\def\sixth{{ 1\over 6}}
\def\third{{1 \over 3}}

\def\twothirds{{2 \over 3}}


\newdimen\pmboffset
\pmboffset 0.022em
\def\oldpmb#1{\setbox0=\hbox{#1}%
 \copy0\kern-\wd0
 \kern\pmboffset\raise 1.732\pmboffset\copy0\kern-\wd0
 \kern\pmboffset\box0}
\def\pmb#1{\mathchoice{\oldpmb{$\displaystyle#1$}}{\oldpmb{$\textstyle#1$}}
      {\oldpmb{$\scriptstyle#1$}}{\oldpmb{$\scriptscriptstyle#1$}}}


%

\nref\GellMann{M. Gell-Mann, P. Ramond and R. Slansky, in {\it
 Supergravity}, edited by F. van Nieuwenhuizen and D. Freedman,
 (North-Holland, Amsterdam, 1979) p. 315\semi
T. Yanagida, Prog. Th. Physics {\bf B135}, 66 (1978).}
\nref\Rajpoot{S. Rajpoot, Phys. Lett. {\bf 191}, 122 (1987).}
\nref\DavidsonI{A. Davidson and K.C. Wali, Phys. Rev. Lett. {\bf 59}, 393
 (1987)\semi
A. Davidson and K.C. Wali, Phys. Rev. Lett. {\bf 60}, 1813 (1988).}
\nref\BabuI{K.S. Babu and R.N. Mohapatra, Phys. Rev. Lett. {\bf 62},
 1079 (1989).}
\nref\BabuII{K.S. Babu and R.N. Mohapatra, Phys. Rev. {\bf D41}, 1286
 (1990).}
\nref\Barr{S.M. Barr, D. Chang and G. Senjanovic, Phys. Rev. Lett. {\bf 67},
 2765 (1991).}
\nref\GG{H. Georgi and S. L. Glashow, Phys. Rev. Lett. {\bf 32}, 438 (1974).}
\nref\Fritzsch{H. Georgi, in {\it Particles and Fields}, edited by
  C.E. Carlson, (A.I.P., New York,1975) p. 575\semi
H. Fritzsch and P. Minkowski, Ann. Phys. {\bf 93}, 193 (1975).}
\nref\PatiSalam{J.C. Pati and A. Salam, Phys. Rev. {\bf D8}, 1240 (1973)\semi
 J.C. Pati and A. Salam, Phys. Rev. {\bf D10}, 275 (1974).}
\nref\PDB{Review of Particle Properties, Phys. Rev. {\bf D45}, Part 2
 (1992).}
\nref\Giveon{A. Giveon, L.J. Hall and U. Sarid, Phys. Lett. {\bf B271},
 138 (1991).}
\nref\Bethke{S. Bethke, Talk presented at XXVI Int. Conf. High Energy
 Physics, Dallas, Aug. 6-12, 1992.}
\nref\Georgi{H. Georgi, Nucl. Phys. {\bf B156}, 126 (1979).}
\nref\DavidsonII{A. Davidson and K.C. Wali, Phys. Rev. Lett. {\bf 58},
 2623 (1987).}
\nref\Cho{P. Cho and S. Glashow, unpublished.}
\nref\Quinn{H. Georgi, H.R. Quinn and S. Weinberg, Phys. Rev. Lett. {\bf 33},
 451 (1974).}
\nref\Dawson{S. Dawson and H. Georgi, Phys. Rev. Lett. {\bf 43},
 821 (1979).}
\nref\GeorgiII{H. Georgi, Private communication.}
\nref\Beall{G. Beall, M. Bander and A. Soni, Phys. Rev. Lett. {\bf 48},
 848 (1982).}
\nref\Ecker{G. Ecker and W. Grimus, Nucl. Phys. {\bf B258}, 328 (1985).}
\nref\Deshpande{N.G. Deshpande and R.J. Johnson, Phys. Rev. {\bf D27},
 1193 (1983).}
\nref\DeRujula{A. DeRujula, H. Georgi and S.L. Glashow, in {\it Fifth
 Workshop on Grand Unification}, edited by K. Kung, H. Fried and P. Frampton
 (World Scientific, Singapore, 1984) p. 88.}
\nref\Carlson{E.D. Carlson and M.Y. Wang, HUTP-92/A057 (1992).}


%
\nfig\sinsqfig{Evolution of $\sinsq(\u)$ over the range $\Mz\le\u\le\Mgut$
in the $SU(4)^3$ model. Dashed lines mark the locations of the
intermediate $\vR$, $\LLR$ and $\LC = \Lambda_\R$ scales.}
\nfig\pdecay{Dominant contribution to proton decay from $\chi$ scalar
exchange in the $SU(4)^3$ model.  Primed and unprimed fields denote mass
and gauge eigenstates respectively.}


\def\CITTitle#1#2{\nopagenumbers\abstractfont
\hsize=\hstitle\rightline{#1}
\vskip 1in\centerline{\titlefont #2~\footnote{$\hskip-0.2em^*\hskip-0.5em$}
  {\vskip-0.26in{\sevenrm Work supported in part by the U.S. Dept. of
  Energy under Contract no. DEAC-03-81ER40050.}}}
 \abstractfont\vskip .5in\pageno=0}

\CITTitle{{\baselineskip=12pt plus 1pt minus 1pt
  \vbox{\hbox{CALT-68-1859}\hbox{DOE RESEARCH AND}\hbox{DEVELOPMENT
  REPORT}}}}{Unified Universal Seesaw Models}
\centerline{Peter Cho}
\bigskip\centerline{\it California Institute of Technology, Pasadena, CA
  91125}

\vskip .6in
\centerline{\bf Abstract}
\bigskip

	A set of Grand Unified Theories based upon the gauge groups
$SU(5)_\L \times SU(5)_\R$, $SO(10)_\L \times SO(10)_\R$ and
$SU(4)_\C \times SU(4)_\L \times SU(4)_\R$ is explored.  Several novel
features distinguish these theories from the well-known $SU(5)$, $SO(10)$
and $SU(4)_\C \times SU(2)_\L \times SU(2)_\R$ models which they
generalize.  Firstly, Standard Model quarks and leptons are accompanied by
and mix with heavy $SU(2)_\L \times SU(2)_\R$ singlet partners.
The resulting fermion mass matrices are seesaw in form.  Discrete parity
symmetries render the determinants of these mass matrices real and
eliminate CP violating gauge terms.  The unified seesaw models consequently
provide a possible resolution to the strong CP problem.  Secondly, $\sinsq$
at the unification scale is numerically smaller than the experimentally
measured $Z$ scale value.   The weak angle must therefore increase as it
evolves down in energy.  Finally, proton decay is suppressed by small
seesaw mixing factors in all these theories.

\Date{4/93}

\newsec{Introduction}

	Among the many questions left unanswered by the Standard Model of
particle physics, the origin of fermion masses ranks as one of the most
intriguing and important.  Details of the fermion mass spectrum remain a
perplexing mystery, and even its gross features are not understood.  One
general characteristic which remains unexplained within the context of the
minimal Standard Model is the disparity between the electroweak scale and
quark and lepton masses.  In the most extreme case, the mass of the electron
is roughly a million times smaller than the weak scale.  This dichotomy can
of course be accommodated in the Standard Model by tuning certain Yukawa
couplings to be sufficiently small.  However, a more natural explanation for
this mass gap would be preferable.

	In the past few years, such an explanation has been offered in which
the familiar neutrino seesaw mechanism \GellMann\ is applied to charged
fermions as well \refs{\Rajpoot,\DavidsonI}.  This universal seesaw proposal
necessitates the introduction of new heavy partners for each of
the known Standard Model fermions with which they mix.
The lightness of observed quarks and leptons then results as a natural
consequence of the seesaw mechanism.  This scheme obviously works best for
the first generation of fermions and worst for the third.  In particular,
achieving the anomalously large mass for the top quark is problematic.
Nonetheless, the basic idea of a universal seesaw mechanism is appealing and
sheds some light on the fermion mass puzzle.

	A second fundamental question left unaddressed by the Standard
Model but which theories with a universal seesaw mechanism can resolve is
the strong CP problem.  Such theories generally possess a parity symmetry
which prohibits a CP violating $\th_\QCD$ term from appearing in the QCD
Lagrangian and renders Yukawa coupling matrices hermitian.
So while the fermion mass matrix can be complex and generate weak CP
violation, the argument $\th_\QFD$ of its determinant is zero.  The
physically observable parameter $\bar{\th}=\th_\QCD+\th_\QFD$ consequently
vanishes at tree order.  Universal seesaw models thus offer a solution to
the strong CP problem which does not involve axions
\refs{\BabuI,\BabuII,\Barr}.

	The universal seesaw mechanism has been studied in the past
mainly within the context of the left-right symmetric
$SU(3)_\C \times SU(2)_\L \times SU(2)_\R \times U(1)$ model.
In this paper, we explore a number of possibilities for embedding this
mechanism within a unified theory.   In particular, we investigate
models based upon the gauge groups $SU(5)_\L \times SU(5)_\R$,
$SO(10)_\L \times SO(10)_\R$ and $SU(4)_C \times SU(4)_\L \times SU(4)_\R$.
As we shall see, such unified theories provide a rationale for the
seemingly ad hoc introduction of heavy $SU(2)_\L \times SU(2)_\R$ singlet
fermions in their un-unified counterparts.  Moreover, these particular
models generalize the well-known $SU(5)$ \GG\ and $SO(10)$ \Fritzsch\
Grand Unified Theories and the $SU(4)_\C \times SU(2)_\L \times SU(2)_\R$
Pati-Salam model \PatiSalam.  So they are of interest in their own right.

	To help guide our exploration, we will adopt the following set of
unified seesaw model building rules:

\item{I.} The model must reproduce the measured $Z$ scale values for the
Standard Model couplings \refs{\PDB,\Giveon,\Bethke}
\eqna\Zscaleparams
$$ \eqalignno{ \sinsq (\Mz) &= 0.2325 \pm 0.0008 & \Zscaleparams a \cr
\aEM^{-1} (\Mz) &= 127.8 \pm 0.2 & \Zscaleparams b \cr
\aS (\Mz) &= 0.118 \pm 0.007. & \Zscaleparams c \cr} $$

\item{II.} The model must satisfy other phenomenological constraints such
as limits on new particle masses and bounds on proton decay.

\item{III.} The model should incorporate heavy $SU(2)_\L \times SU(2)_\R$
singlet fermions which mix with Standard Model quarks and leptons to allow
for a seesaw mass matrix whose determinant is real.

\item{IV.} The model should contain fermions in anomaly free but complex
representations in accordance with the ``survival hypothesis'' \Georgi.

\item{V.} The model preferably maintains left-right symmetry from the
unification scale down to the Standard Model subgroup level.

\medskip\noindent
These requirements are listed in approximate order of importance.   The
first two experimental constraints are binding and must be satisfied by any
realistic Grand Unified Theory.  The remaining theoretical guidelines are
more negotiable.  In particular, the last item is included only to help
restrict the large number of possible symmetry breaking patterns in the
models we shall explore.  So we may relax this final aesthetic condition
in order to fulfill the other more stringent requirements in this list.

	The remainder of our paper is organized as follows.  We present the
$SU(5) \times SU(5)$ and $SO(10) \times SO(10)$ models in sections 2 and 3.
These theories illustrate the basic features of all unified universal
seesaw models.  They also serve as warmups for the $SU(4) \times SU(4)
\times SU(4)$ model which is discussed in greater detail in section 4.
{}Finally, we close in section 5 with some indications for possible further
investigation of this new class of Grand Unified Theories.

\newsec{The prototype $\pmb{SU(5)\times SU(5)}$ model}

	The first model that we shall explore is based upon the gauge
group $G=SU(5)_\L \times SU(5)_\R$.  This theory represents an obvious
generalization of the Georgi-Glashow $SU(5)$ model \GG\ and shares
many of its attractive features.  It is also the simplest unified seesaw
model and has been analyzed in the past \refs{\DavidsonII,\Cho}.  While
this theory ultimately turns out not to be phenomenologically viable, it
is worth reviewing since many of its basic characteristics are common to
all unified universal seesaw models.

	To begin, we impose a $Z_2$ symmetry on the chiral theory
which combines a spatial inversion with interchanging the $SU(5)$ factors
in the product group $G$.  Such a discrete symmetry is needed to ensure the
equality of the $SU(5)_\L$ and $SU(5)_\R$ couplings constants above the
unification sale.  In its absence, the couplings would run differently and
diverge even if they were set equal at one particular renormalization
point.  The generalized parity operation enforces a left-right symmetry on
the Lagrangian which may be violated only softly by superrenormalizable
interactions.  It also dictates a one-to-one correspondence among matter
field representations of $SU(5)_\L$ and $SU(5)_\R$.  The spectrum of this
theory consequently exhibits an explicit parity doubling.

	We next embed the Standard Model within the Grand Unified Theory
following the Georgi-Glashow model blueprint.  Color
$SU(3)$ and weak $SU(2)$ are identified with the diagonal $SU(3)_{\L+\R}$
subgroup of $G$ and the $SU(2)_\L$ subgroup of $SU(5)_\L$
respectively.  $U(1)_\EM$ is generated by the diagonal sum of the familiar
$SU(5)_\L$ and $SU(5)_\R$ electric charge generators.  The
$SU(3) \times SU(2) \times U(1)$ content of a single fermion family
representation
\eqn\familyrepI{\CF \sim (\bar{5}+10,1)+(1,\bar{5}+10)}
is then readily established.  The fermions' colors, flavors and electric
charges are indicated by conventional letter names in the matrices below:
$$ (\psi_\L)_{i\,} \, = \pmatrix{D_1^c \cr D_2^c \cr D_3^c \cr e \cr
-\nu \cr}_\L \sim (\bar{5},1) \qquad
\> (\Psi_\L)^{i\,j\,} = {1\over \sqrt{2}}
\pmatrix{ 0 & U_3^c & - U_2^c & -u_1 & -d_1 \cr
         -U_3^c & 0 & U_1^c   & -u_2 & -d_2 \cr
	 U_2^c & -U_1^c & 0   & -u_3 & -d_3 \cr
	 u_1 & u_2 & u_3 & 0 & -E^c \cr
	 d_1 & d_2 & d_3 & E^c & 0 \cr}_\L \sim (10,1) $$
\eqn\family{\>}
$$ (\psi_\R)_{i'} = \pmatrix{D_1^c \cr D_2^c \cr D_3^c \cr e \cr
-\nu \cr}_\R \sim (1,\bar{5}) \quad\;\;
(\Psi_\R)^{i'j'} = {1\over \sqrt{2}}
\pmatrix{ 0 & U_3^c & - U_2^c & -u_1 & -d_1 \cr
         -U_3^c & 0 & U_1^c   & -u_2 & -d_2 \cr
	 U_2^c & -U_1^c & 0   & -u_3 & -d_3 \cr
	 u_1 & u_2 & u_3 & 0 & -E^c \cr
	 d_1 & d_2 & d_3 & E^c & 0 \cr}_\R \sim (1,10). $$
Three generations of families are assumed as in the Standard Model and
assigned to three copies of $\CF$.

	We now specify a simple symmetry breaking pattern that starts with
the unified chiral gauge group and cascades down to unbroken color and
electromagnetism:
\eqn\patternI{\eqalign{SU(5)_\L &\times SU(5)_\R \cr
& \downarrow \Mgut \cr
SU(3)_\L \times SU(2)_\L \times U(1)_\L & \times SU(3)_\R \times
SU(2)_\R \times U(1)_\R \cr
& \downarrow \LLR \cr
SU(3)_{\L+\R} \times SU(2)_\L & \times SU(2)_\R \times U(1)_{\L+\R} \cr
& \downarrow \vR \cr
SU(3)_{\L+\R} \times SU&(2)_\L \times U(1)_\Y \cr
& \downarrow \vL \cr
SU(3)_{\L+\R} & \times U(1)_{\EM} .  \cr } }
A minimal number of fundamental Higgs fields is introduced into the
theory to achieve this pattern.  As in the Georgi-Glashow model,
$SU(5)_\L$ and $SU(5)_\R$ are broken with scalars $\PhiL\sim (24,1)$ and
$\PhiR\sim (1,24)$ that transform in their adjoint representations.  The
fermion families decompose under the resulting
$(SU(3) \times SU(2) \times U(1))^2$ subgroup as
\eqn\familydecomp{\eqalign{
\CF & \sim \bigl[ (\bar{3},1,1,1)^{\third,0} +
(1,\bar{2},1,1)^{-{1\over 2},0}+
(\bar{3},1,1,1)^{-\twothirds,0} + (3,2,1,1)^{\sixth,0} +
(1,1,1,1)^{1,0} \bigr]_\L \cr
& + \bigl[ (1,1,\bar{3},1)^{0,\third} + (1,1,1,\bar{2})^{0,-{1\over 2}} +
(1,1,\bar{3},1)^{0,-\twothirds} +
(1,1,3,2)^{0,\sixth} + (1,1,1,1)^{0,1} \bigr]_\R . \cr} }
The subsequent breaking of chiral color and chiral hypercharge to
their diagonal subgroups is performed at the $\LLR$ scale by Higgs fields
$\omega\sim ({\overline 5},5)$ and $\Omega\sim (10,\overline{10})$.
If these scalars develop the vacuum expectation values
\eqn\omegavevs{
\vev{\omega}^{1'}_1 = \vev{\omega}^{2'}_2 = \vev{\omega}^{3'}_3 =
\vev{\Omega}^{[12]}_{[1'2']} = \vev{\Omega}^{[23]}_{[2'3']} =
\vev{\Omega}^{[31]}_{[3'1']} = \vev{\Omega}^{[45]}_{[4'5']} = \LLR,}
the chirally colored $(\bar{3},1,1,1)$ and $(1,1,\bar{3},1)$ and
chirally hypercharged $(1,1,1,1)^{1,0}$ and $(1,1,1,1)^{0,1}$ fields in
\familydecomp\ marry together and acquire Dirac masses
through the Yukawa interactions
\eqna\Yukawa
$$ \eqalignno{{\cal L}_{\rm Yukawa}(\omega,\Omega) &=
-\Bigl[f_\omega ({\overline \psiL})^i (\omega)^{i'}_i (\psiR)_{i'} +
{f_\Omega\over 2} ({\overline \PsiL})_{ij} (\Omega)^{ij}_{i'j'}
(\PsiR)^{i'j'} \bigr] + \hc & \Yukawa a \cr} $$
These fourteen $SU(2)_\L \times SU(2)_\R$ singlet fermions automatically
emerge in the unified theory as the heavy seesaw partners that are added
by hand in un-unified seesaw models.  They are denoted by capital letters
in \family.  The remaining sixteen fields in \familydecomp\ reside
within $SU(2)$ doublets and stay massless at the $\LLR$ scale.  They
essentially correspond to the known Standard Model fermions plus a right
handed neutrino and are represented by the lower case letters in \family.

	The last two steps in pattern \patternI\ are accomplished
by scalars $\phiL\sim (5,1)$ and $\phiR\sim (1,5)$ which break
$SU(2)_\L$ and $SU(2)_\R$ via the VEV's $\vev{\phi_{\L,\R}} =
(0,0,0,0,v_{\L,\R}/\sqrt{2})^\T$.  Masses connecting heavy and light
fermions are then generated by the Yukawa terms
$$ \eqalignno{{\cal L}_{\rm Yukawa}(\phi) &=
f_\phi \Bigl[ (\psiL^\T)_i C (\PsiL)^{ij} ({\phiL}^\dagger)_j
+ (\psiR^\T)_{i'} C (\PsiR)^{i'j'} ({\phiR}^\dagger)_{j'} \Bigr]
& \cr
&+ f'_\phi \Bigl[ \epsilon_{ijklm} (\PsiL^\T)^{ij} C
(\PsiL)^{kl} (\phiL)^m +\epsilon_{i'j'k'l'm'} (\PsiR^\T)^{i'j'} C
(\PsiR)^{k'l'} (\phiR)^{m'} \Bigr] + \hc & \cr
&& \Yukawa b \cr}  $$
The quark and lepton mass matrices thus assume the seesaw forms
\eqn\Lfermi{\eqalign{{\cal L}_{\rm mass} &=
\pmatrix{ \bar{u_\L} & \bar{U_\L} \cr}
\pmatrix{ 0 & \sqrt 2 {f'_\phi}^\dagger \vL \cr
 \sqrt 2 f'_\phi \vR & f_\Omega^\T \LLR \cr}
\pmatrix{ u_\R \cr U_\R \cr} \cr\cr
& \quad + \pmatrix{ \bar{d_\L} & \bar{D_\L} \cr}
\pmatrix{ 0 & \half f_\phi^\dagger \vL \cr
 \half f_\phi \vR & f_\omega^\T \LLR \cr}
\pmatrix{d_\R \cr D_\R \cr} \cr\cr
& \quad + \pmatrix{ \bar{e^+_\R} & \bar{E^+_\R} \cr}
\pmatrix{ 0 & \half f_\phi \vL \cr
 \half f_\phi^\dagger \vR & f_\Omega \LLR \cr}
\pmatrix{e^+_\L \cr E^+_\L \cr} + {\rm h.c.} \cr} }

	It is important to recall that the fermion fields are
$\NF=3$ dimensional vectors in family space.  The Yukawa
couplings in eqns.~\Yukawa{a,b}\ are consequently $\NF \times \NF$ matrices
with generation indices that have been suppressed.
Parity constrains $f_\omega$ and $f_\Omega$ to be hermitian, while
the form of the second term in \Yukawa{b}\ automatically renders $f'_\phi$
symmetric.  If these Yukawa couplings are approximately comparable in
magnitude, then the mass matrices have the well-known seesaw eigenvalues
\eqn\evalues{\eqalign{
m & \simeq - O(f {\vL\vR \over \LLR}) \cr
M & \simeq O(f \LLR) \cr}}
and corresponding eigenvectors
\eqn\evectors{\eqalign{
\pmatrix{q' \cr  Q' \cr} = \pmatrix{ 1 & -O(\vR/\LLR) \cr
O(\vL/\LLR) & 1 \cr}
\pmatrix{q \cr Q \cr} }}
provided $\vL\vR \ll \LLR^2$.  We thus recover the universal seesaw
mechanism in this $SU(5) \times SU(5)$ theory.

	The fermion mass matrices in \Lfermi\ are generally complex
and induce weak CP violation as in the Standard Model.  But their
determinants are real.   This can be simply verified by rewriting the
down-type quark matrix for example as
\eqn\MdD{\CM_{dD} = \pmatrix{1 & 0 \cr 0 & \vR/\LLR \cr}
\pmatrix{ 0 & \half f^\dagger_\phi \LLR \cr
 \half f_\phi \LLR & f_\omega^\T \LLR^3/\vL\vR \cr}
\pmatrix{1 & 0 \cr 0 & \vL/\LLR}.}
Since the diagonal matrices are real while the middle matrix is hermitian,
we conclude that ${\rm arg} ({\rm det} \CM_{dD}) = 0$.  So as a result of
the generalized parity symmetry in the $SU(5) \times SU(5)$ model, the
complex argument $\th_\QFD$ of the total mass matrix as well as the
$\th_{\QCD}$ term in the QCD Lagrangian vanish at tree order. The seesaw
GUT therefore provides a possible solution to the strong CP problem.

	Unfortunately, the symmetry breaking pattern in \patternI\ is
not phenomenologically viable.  Recall that once the embedding of the
electroweak subgroup inside the gauge group $G$
is specified, the value of $\sinsq$ at the unification scale is fixed:
\eqn\sinsqGUT{\sinsq(\Mgut) = {\Tr (T^3_\L)^2 \over \Tr Q^2} = {3 \over
16} = 0.1875 .}
In this $SU(5)\times SU(5)$ model, there are twice as many electrically
charged fermions as in the $SU(5)$ theory but precisely the same number
of weak $SU(2)_\L$ doublets.  So $\sinsq(\Mgut)$ is half as large as in
the Georgi-Glashow model \GG\ and starts out numerically smaller
than $\sinsq(\Mz)=0.2325$.  Moreover, renormalization effects decrease the
value of $\sinsq(\u)$ for $\u < \Mgut$ in the $SU(5) \times SU(5)$ theory
just as in the $SU(5)$ model \Quinn.  Therefore, pattern \patternI\ cannot
duplicate the $Z$ scale measurement and must be rejected.

	One can try to search for alternate breaking patterns in
which $\sinsq$ increases as it evolves down in energy from the GUT
scale.  Maximal enhancement is achieved if the first stage of symmetry
breaking is taken to be $SU(5)_\L \times SU(5)_\R \to SU(3)_\L \times
SU(2)_\L \times U(1)_\L \times SU(5)_\R$ \refs{\DavidsonII,\Cho}.  This
clearly leads to trouble with proton decay.  Moreover,
detailed calculation demonstrates that this asymmetrical breaking pattern
still cannot yield the values for the Standard Model couplings in
\Zscaleparams{}\ \Cho.  We therefore conclude that an $SU(5) \times SU(5)$
seesaw theory is ruled out.

\newsec{The $\pmb{SO(10) \times SO(10)}$ model}

	The GUT scale value for $\sinsq$ tends to be small in all unified
universal seesaw models as we have seen in the particular case of
$SU(5) \times SU(5)$.   So in order for these theories to
have any chance of being phenomenologically viable, we must find some
mechanism for enhancing $\sinsq$ as it evolves down in energy from the
unification scale.   We will illustrate a general strategy
for overcoming this problem in the context of an $SO(10) \times SO(10)$
model.

	This second theory represents an obvious generalization of the
first considered in the preceding section, and a number of parallel
features can immediately be established.  For instance, a discrete
interchange symmetry must again be imposed on the separate factors in
the gauge group $G = SO(10)_\L \times SO(10)_\R$. As a result, particle
representations occur in pairs, and fermion families in particular
transform as
\eqn\familyrepII{\CF \sim (16,1)+(1,16)}
which generalizes the $SU(5) \times SU(5)$ assignments in \familyrepI.
There are however some significant differences between the two models.
Most importantly, the larger size of $SO(10) \times SO(10)$ allows several
new possibilities for electroweak subgroup embedding and symmetry breaking.
As we shall see, this greater flexibility provides the key to
increasing $\sinsq$ at the $Z$ scale.

	Among the many different potential breaking schemes, we focus upon
the following pattern which maintains explicit left-right symmetry down
to the Standard Model:
\vfill\eject
\eqn\patternII{\eqalign{
SO(10)_\L &\times SO(10)_\R \cr
L^\a \sp\quad & \sp\qquad R^\a \cr
\gtenL \sp\quad & \sp\sp\quad \gtenR \cr
& \downarrow \Mgut \cr
SU(4)_\L \times SU(2)_\L \times SU(2)'_\L & \times SU(4)_\R \times
 SU(2)_\R \times SU(2)'_\R \cr
U^\A_\L \quad\qquad T^i_\L \quad\qquad {T^i_\L}' \sp\quad & \qquad
U^\A_\R \quad\qquad T^i_\R \quad\qquad {T^i_\R}' \cr
\gfourL \quad\qquad \gtwoL \quad\qquad
 \gtwoLp \sp\quad & \qquad \gfourR  \quad\qquad \gtwoR \quad\qquad \gtwoRp
\cr
& \downarrow \LLR \cr
SU(4)_{\L+\R} \times SU(2)_\L & \times SU(2)_\R \times SU(2)'_{\L+\R} \cr
U^\A=U^\A_\L+U^\A_\R \sp\quad T^i_\L \quad & \qquad T^i_\R \qquad
 {T^i}'={T^i_\L}'+{T^i_\R}' \cr
\gfour \qquad\qquad \gtwoL \quad &
 \qquad \gtwoR \sp\quad\qquad \gtwop \cr
& \downarrow \LC \cr
SU(3)_{\L+\R} \times SU(2)_\L & \times SU(2)_\R \times U(1)_{\L+\R} \cr
U^a \qquad\qquad T^i_\L \quad & \qquad T^i_\R \quad S={T^3}'
 +\sqrt\twothirds U^{15} \cr
\gthree \qquad\qquad \gtwoL \quad &
 \qquad \gtwoR \sp\quad\qquad \gone \cr
& \downarrow \vR \cr
SU(3)_{\L+\R} \times SU&(2)_\L \times U(1)_\Y \cr
U^a \qquad\qquad & T^i_\L \sp\quad Y/2=T^3_\R+S \cr
g_3 \qquad\qquad & \gtwoL \sp\qquad g' \cr
& \downarrow \vL \cr
SU(3)_{\L+\R} & \times U(1)_{\EM} \cr
U^a \sp\quad & \quad Q=T^3_\L+Y/2 \cr
g_3 \>\sp\quad & \qquad\sp e \cr
}}
We have listed underneath each of the subgroup factors in this pattern
our nomenclature conventions for the associated generators and coupling
constants.
\foot{The ranges of the $SO(10)$, $SU(4)$, $SU(3)$ and $SU(2)$ generator
labels $\a$, $A$, $a$ and $i$ are respectively $1 \dash 45$, $1 \dash 15$,
$1 \dash 8$ and $1 \dash 3$.}

	The generators at each level in \patternII\ are linear combinations
$H=\sum_i c_i G_i$ of those at the previous level, and the
corresponding couplings are related as $h^{-2}=\sum_i (g_i/c_i)^{-2}$.  In
particular, the electric charge generator
\eqn\QII{Q=T^3_\L + T^3_\R + {T^3_\L}'+{T^3_\R}'+\sqrt\twothirds U^{15}_\L
+\sqrt\twothirds U^{15}_\R}
of the final unbroken $U(1)_\EM$ subgroup is a combination of
elements in the Cartan subalgebras of $SU(2)_{\scriptscriptstyle L,R}$,
$SU(2)'_{\scriptscriptstyle L,R}$ and $SU(4)_{\scriptscriptstyle L,R}$.
The corresponding relation among these groups' coupling constants
\eqn\couplingreln{e(\u)^{-2} = \gtwoL (\u)^{-2} + \gtwoR (\u)^{-2}
+ \gtwoLp (\u)^{-2} + \gtwoRp (\u)^{-2}
+\twothirds \gfourL (\u)^{-2} +\twothirds \gfourR (\u)^{-2}}
fixes the GUT scale value of the weak mixing angle:
\eqn\sinsqGUT{\sinsq(\Mgut) = {e(\Mgut)^2 \over \gtwoL (\Mgut)^2}
 = {3 \over 16} = 0.1875.}

	Since the $SU(3)\times SU(2) \times U(1)$ content of the
fermion representation \familyrepII\ in the $SO(10) \times SO(10)$ model
is identical to that of \familyrepI\ in the $SU(5) \times SU(5)$ theory
except for an additional electrically neutral $SU(2)_\L \times SU(2)_\R$
singlet field, we have again found a value for $\sinsq (\Mgut)$ which is
precisely half as large as in the Georgi-Glashow model.  But the behavior
of $\sinsq$ below the unification scale is qualitatively different:
\eqna\sinsqcases
$$ \eqalignno{\sinsq(\u)=\cases{
\displaystyle{1 \over 1+\Bigl({\gtwoL\over\gtwoR}\Bigr)^2
 + \Bigl({\gtwoL\over\gtwoLp} \Bigr)^2 + \Bigl({\gtwoL\over\gtwoRp}\Bigr)^2
 + \twothirds\Bigl({\gtwoL\over\gfourL}\Bigr)^2
 + \twothirds\Bigl({\gtwoL\over\gfourR}\Bigr)^2} & $\LLR\le\u\le\Mgut$
 \sinsqcases a \cr &\cr&\cr
\displaystyle{1 \over 1+\Bigl({\gtwoL\over\gtwoR}\Bigr)^2
 + \Bigl({\gtwoL\over\gtwop} \Bigr)^2
 + \twothirds\Bigl({\gtwoL\over\gfour}\Bigr)^2} & $\LC\le\u\le\LLR\sp$
 \quad \sinsqcases b \cr &\cr&\cr
\displaystyle{1 \over 1+\Bigl({\gtwoL\over\gtwoR}\Bigr)^2
 + \Bigl({\gtwoL\over\gone} \Bigr)^2} & $\vR\le\u\le\LC$
 \qquad \sinsqcases c \cr &\cr&\cr
\displaystyle{1 \over 1+\Bigl({\gtwoL\over g'} \Bigr)^2}
 & $\vL\le\u\le\vR\sp$ \qquad \sinsqcases d \cr}} $$
In \sinsqcases{a}, the $SU(2)$ couplings $\gtwoL$, $\gtwoR$, $\gtwoLp$ and
$\gtwoRp$ are all asymptotically free and increase as they run down
in energy.  However, the $SU(4)$ couplings $\gfourL$ and
$\gfourR$ increase even faster.  So the denominator in \sinsqcases{a}\
decreases and the total fraction grows larger for $\u < \Mgut$.
This rising trend continues until the $\LC$ scale is reached.
At that point, $\sinsq(\u)$ begins to decrease and continues downward
all the way to $\u = \Mz$.  The final sign and magnitude of the net change
in $\sinsq$ depend in detail upon the numerical values of the various
intermediate scales and beta functions of the couplings appearing within
the multilevel pattern \patternII.  But we at least see how an enhancement
of the weak mixing angle may be achieved in principle
\refs{\Dawson,\GeorgiII}.

	Unification by itself cannot uniquely determine all the
symmetry breaking scales in \patternII.  However, a number of
phenomenological considerations restrict their values.   Firstly,
$K\dash\bar{K}$ mixing places lower mass limits of $1.6\dash 2.5 \TeV$ on
$W_\R^\pm$ gauge bosons in manifestly left-right symmetric
$SU(2)_\L \times SU(2)_\R \times U(1)$
theories \refs{\Beall,\Ecker}.  Therefore, $\vR$ must lie at least in
the multi-TeV region.  Secondly, limits on the lepton family number
violating decay $K_\L \to \mu^\pm e^\mp$ provide a bound on the $\LC$
scale, for it is mediated by $SU(4)_{\L+\R}$ gauge boson exchange.
Including renormalization effects \Deshpande, we estimate that the
branching fraction limit \PDB
\eqn\brfrac{
{\Gamma(K_\L \to \mu^\pm e^\mp) \over \Gamma(K^+ \to \mu^+\nu_\mu)}
< 3.54 \times 10^{-11}}
restricts $\LC \gtap 10^6 \GeV$.  Finally, the unification scale $\Mgut$
must be sufficiently large to allow for an acceptable proton lifetime.

	It is useful to imagine constructing a low energy effective field
theory at each symmetry breaking stage in pattern \patternII\ in order to
simplify the renormalization group analysis of coupling constant evolution.
Particles that can acquire masses at a certain scale are integrated out
together and do not contribute to subsequent renormalization group running.
We thus find the following one-loop gauge boson and fermion contributions
to the $U(1)$ and $SU(n)$ beta functions $\b(g_n)=b_n g_n^3/
16\pi^2$:
\foot{Ordinary quarks and leptons are singlets under $SU(2)'_{\L+\R}$,
while their seesaw partners acquire heavy masses and decouple at the $\LLR$
scale in \patternII.  There is consequently no fermion contribution to the
$SU(2)'_{\L+\R}$ beta function coefficient $b'_2$.}
\eqn\betacoeffs{\eqalign{b_\Y &= {20 \over 3} \NF \cr
b_1 &= {8 \over 9} \NF \cr
b'_2 &= -{22 \over 3} \cr
b_n &=- \Bigl( {11 n \over 3} - \fourthirds \NF \Bigr). \cr}}
It is then straightforward to integrate the renormalization group equations
to obtain a linear system of equations that relates the three high energy
quantities $\aten(\Mgut)=g_{\scriptscriptstyle 10}(\Mgut)^2 /4\pi$,
$\log (\Mgut/\LLR)$ and $\log( \LLR/\LC)$ to the three
low energy parameters $\sinsq(\Mz)$, $\aEM(\Mz)$ and $\aS(\Mz)$:
\eqn\system{\eqalign{& \pmatrix{
{13 \over 3} & 3 \btwoL + \fourthirds\bfourL & \btwoL+\btwop+\twothirds
 \bfour \cr
1 & \btwoL & \btwoL \cr
2 & 2 \bfourL & \bfour \cr}
\pmatrix{
{2\pi / \aten(\Mgut)} \cr
\log{\Mgut/\LLR} \cr
\log{\LLR/\LC} \cr} = \cr
& \qquad\quad\qquad \pmatrix{
{2\pi\cossq(\Mz)/\aEM(\Mz)}-(\bone+\btwo)\log{\LC/\vR}-b_\Y \log{\vR/\Mz} \cr
{2\pi\sinsq(\Mz)/\aEM(\Mz)}-\btwoL \log{\LLR/\Mz} \cr
{2\pi/\aS(\Mz)}-\bthree \log{\LC/\Mz} \cr}. \cr}}
Unfortunately, no consistent solution to this matrix equation
exists which satisfies the phenomenological restrictions on the intermediate
scales and reproduces the high precision numbers in \Zscaleparams{}.
A fit for the GUT scale parameters based upon the inputs $\vR = 5 \TeV$,
$\LC=1000 \TeV$ and $\LLR = 100,000 \TeV$ yields the results
\eqn\GUTparams{\eqalign{\aten(\Mgut) &= 0.025 \cr
\Mgut &= 1.3 \times 10^{15} \GeV \cr}}
which imply the $Z$ scale values
\eqn\postdictions{\eqalign{\sinsq(\Mz) &= 0.197 \cr
\aEM^{-1}(\Mz) &= 124.3 \cr
\aS(\Mz) &= 0.137. \cr}}
The match between these theoretical numbers and the experimental
measurements in \Zscaleparams{}\ is obviously poor.  Nonetheless, we see
that the basic strategy of embedding part of the hypercharge generator
within an asymptotically free subgroup has led to an increase in
$\sinsq(\Mz)$ over its unification value \GeorgiII.  This trick must
generally be employed in any unified universal seesaw model.

	At this point, we could explore other symmetry breaking schemes for
the $SO(10)\times SO(10)$ theory in which manifest left-right symmetry is
broken at an earlier stage than in pattern \patternII\ so as to further
enhance the value for $\sinsq(\Mz)$.  Alternatively, we could continue to
search for a phenomenologically viable chiral $G_\L \times G_\R$ model based
upon an even larger group such as $E_6 \times E_6$.  But we turn instead to
explore a somewhat different theory with the gauge structure
$G_\C \times G_\L \times G_\R$ in the following section.

\newsec{The $\pmb{SU(4) \times SU(4) \times SU(4)}$ model}

	The prototypical example of a unified $G_\C \times G_\L \times G_\R$
theory is the $SU(3)_\C \times SU(3)_\L \times SU(3)_\R$ model \DeRujula.
This amusing ``trinification'' theory has been studied in the past as an
alternative to $SU(5)$ and $SO(10)$ unification.  The $SU(3)^3$ model
however cannot accommodate a heavy $SU(2)_\L \times SU(2)_\R$ partner
for each Standard Model fermion.  So we are led to consider
the next simplest possibility based upon the gauge group
\eqn\gaugegroup{G = SU(4)_\C \times SU(4)_\L \times SU(4)_\R}
which is supplemented with a cyclic $Z_3$ symmetry to ensure equality among
the separate $SU(4)$ coupling constants.   This theory represents an obvious
generalization of $SU(3)^3$ trinification as well as the Pati-Salam model
\PatiSalam.  Indeed, Pati and Salam originally proposed $G$ as a possible
global symmetry of nature in which lepton number plays the role of a fourth
color.  The similarities and differences between our model and these others
that have been studied in the past will become evident as we proceed.

	Embedding the Standard Model subgroup within $SU(4)^3$
is straightforward.  We take a generalized set of Gell-Mann matrices
as generators of $SU(4)_\C$.  The first eight members of this set
are associated with color $SU(3)$, while the fifteenth matrix
\eqn\fifteenthgen{
U_\C^{15} = \sqrt{3 \over 2} \pmatrix{\sixth&&& \cr &\sixth&& \cr &&\sixth&
\cr &&&-\half \cr}}
generates a commuting $U(1)_\C$ factor.  For $SU(4)_{\L,\R}$, we use the
set of $4 \times 4$ Pauli matrices
\eqn\Paulis{{\sigma^i \over 2\sqrt{2}},\qquad {\tau^j \over 2\sqrt{2}},
\qquad {\sigma^i \tau^j \over 2\sqrt{2}} \qquad\qquad i,j=1,2,3}
as normalized generators.  The linear combinations
\eqn\LRgens{\eqalign{T^i_{\L,\R} &= {\sigma_i(1+\tau^3) \over 4}
 = \half \pmatrix{\sigma^i & \cr & 0 \cr} \cr
{T^i_{\L,\R}}' &= {\sigma^i(1-\tau^3) \over 4}
 = \half \pmatrix{0 & \cr & \sigma^i \cr} \cr
S_{\L,\R} &= \>\>\> {\tau^3 \over 2 \sqrt{2}} \>\>\>
 = {1\over 2 \sqrt{2}} \pmatrix{1 & \cr & -1 \cr} \cr}}
belong to an $SU(2)_{\L,\R} \times SU(2)'_{\L,\R} \times U(1)_{\L,\R}$
subalgebra of $SU(4)_{\L,\R}$.  Weak $SU(2)$ and its right handed analog
are identified as $SU(2)_\L$ and $SU(2)_\R$.  Finally, we choose
\eqn\QIII{
Q = T^3_\L + {T^3_\L}' + T^3_\R + {T^3_\R}' + \sqrt{2\over 3} U^{15}_\C}
as the generator of electromagnetism.  This definition implies
$\sinsq(\Mgut) = 3/14 = 0.2143$.   While this GUT scale value is
still below the $Z$ scale measurement $\sinsq(\Mz)=0.2325$, it is certainly
closer than the corresponding $\sinsq(\Mgut)=3/16=0.1875$ that we found in
the $SU(5)\times SU(5)$ and $SO(10)\times SO(10)$ models.  So we already see
one clear advantage of the $SU(4)^3$ theory over its predecessors.

	Gauge bosons in this model transform according to the 45-dimensional
representation
\eqn\gaugerep{\CG \sim (15,1,1)+(1,15,1)+(1,1,15)}
which automatically remains invariant under cyclic $Z_3$ permutations.  In
the fermion sector, a single family of left handed fields is assigned to the
anomaly free but complex representation
\eqn\fermrep{\CF \sim (4,\bar{4},1)+(1,4,\bar{4})+(\bar{4},1,4).}
One generation of left handed quarks and leptons along with their seesaw
partners fit snugly inside $(4,\bar{4},1)$, while conjugate fields appear
in $(\bar{4},1,4)$.  The remaining $(1,4,\bar{4})$ contains
a new set of leptons.  All these particles' colors, flavors and electric
charges are indicated in the matrices below:
\eqna\familyIII
$$ \eqalignno{\Psi_\CL(4,\bar{4},1) &=
\pmatrix{d_1 & u_1 & D_1 & U_1 \cr
	 d_2 & u_2 & D_2 & U_2 \cr
	 d_3 & u_3 & D_3 & U_3 \cr
	 e   & \v  & E   & N   \cr}_\L & \familyIII a \cr
&&\cr
\Psi_\LR(1,4,\bar{4}) &=
\pmatrix{I^0 &  I^+    & J^0 & J^+     \cr
	 I^- & {I^0}^c & J^- & {J^0}^c \cr
	 K^0 &  K^+    & L^0 & L^+     \cr
	 K^- & {K^0}^c & L^- & {L^0}^c \cr}_\L & \familyIII b \cr
&&\cr
\Psi_\RC(\bar{4},1,4) &=
\pmatrix{d^c_1 & d^c_2 & d^c_3 & e^c \cr
	 u^c_1 & u^c_2 & u^c_3 & \v^c \cr
	 D^c_1 & D^c_2 & D^c_3 & E^c \cr
	 U^c_1 & U^c_2 & U^c_3 & N^c \cr}_\L. & \familyIII c \cr}$$

	There exist a large number of potential symmetry breaking chains
that start from the GUT group and end with the Standard Model.
The simplest schemes which retain manifest left-right symmetry down to the
$SU(3) \times SU(2) \times U(1)$ subgroup do not sufficiently
enhance $\sinsq$ as it runs down in energy to reproduce the measured
$Z$ scale value.  However, if left-right symmetry is broken either
spontaneously or softly at the first stage, then we can find viable
breaking patterns that lead to phenomenologically interesting results.  One
such possibility is the following:
%
%
\eqn\patternIII{\eqalign{SU(4)_\C \times SU&(4)_\L \times SU(4)_\R \cr
& \downarrow \Lambda_\L  = \Mgut \cr
SU(4)_C \times SU(2)_\L \times SU&(2)'_\L \times U(1)_\L \times SU(4)_\R \cr
& \downarrow \Lambda_\R \cr
SU(4)_\C \times SU(2)_\L \times SU(2)'_\L \times U&(1)_\L \times
 SU(2)_\R \times SU(2)'_\R \times U(1)_\R \cr
& \downarrow \LC \cr
SU(3)_\C \times U(1)_\C \times SU(2)_\L \times SU(2)'_\L & \times U(1)_\L
 \times SU(2)_\R \times SU(2)'_\R \times U(1)_R \cr
& \downarrow \LLR \cr
SU(3)_\C \times SU(2)_\L \times SU&(2)_\R \times U(1)'_\LR \times U(1)_\C \cr
& \downarrow \vR \cr
SU(3)_{\L+\R} \times S&U(2)_\L \times U(1)_\Y \cr
& \downarrow \vL \cr
SU(3)_{\L+\R} & \times U(1)_{\EM} .  \cr } }

	The renormalization group analysis of coupling constant running
in this pattern is similar to that described in the preceding section for
the $SO(10) \times SO(10)$ model.  The only qualitatively new feature that
we include in the $SU(4)^3$ analysis is scalar contributions to beta
functions.  These come from the Higgs sector of the theory which we will
discuss in detail shortly.  The results of the renormalization group analysis
yield a wide range of values for the symmetry breaking scales in \patternIII\
that reproduce the Standard Model parameters in \Zscaleparams{}\ and
satisfy all other phenomenological constraints.  For simplicity, we merge
the intermediate $\Lambda_\R$ and $\LC$ thresholds together and quote one
set of representative values for these scales:
\eqn\scales{\eqalign{ \Lambda_\L &= \Mgut = 6.47 \times 10^{11} \GeV \cr
\Lambda_\R &= \LC = 2.07 \times 10^7 \GeV \cr
\LLR &= 1.0 \times 10^5 \GeV \cr
\vR &= 5.0 \times 10^3 \GeV \cr
\vL &= 2.46 \times 10^2 \GeV. \cr}}
The evolution of $\sinsq$ for this choice of scales is illustrated in
\sinsqfig.

	We now consider the minimal Higgs content of the $SU(4)^3$ model
needed to perform the several stages of symmetry breaking in
\patternIII\ and to provide fermion masses.  The first three steps result
from vacuum expectation values of the adjoint fields in
\eqn\adjscalars{\Phi = \Phi_\C(15,1,1) + \Phi_\L(1,15,1) +
\Phi_\R(1,1,15).}
These scalars' VEV's
\eqn\adjvevs{\vev{\Phi_{\L,\R}} = \Lambda_{\L,\R} \pmatrix{1&&& \cr &1&& \cr
  &&-1& \cr &&&-1 \cr} \quad\qquad
\vev{\Phi_\C} = \LC \pmatrix{1&&& \cr &1&& \cr &&1& \cr &&&-3 \cr}}
break the separate $SU(4)$ factors in $G$ as
\eqna\SUfourfactors
$$ \eqalignno{SU(4)_{\L,\R} \;
&{\buildrel \vev{\Phi_{\L,\R}} \over \longrightarrow} \;
 SU(2)_{\L,\R} \times SU(2)'_{\L,\R} \times U(1)_{\L,\R} & \SUfourfactors
a\cr
SU(4)_\C \; &{\buildrel \vev{\Phi_\C} \over \longrightarrow} \;
 SU(3)_\C \times U(1)_\C. & \SUfourfactors b\cr} $$

	The $SU(2)'_\L$ and $SU(2)'_\R$ subgroups under which the
seesaw fermions transform are subsequently reduced
at the $\LLR$ scale to the diagonal $U(1)'_\LR$ generated by
$S'_\LR={T^3_\L}'+{T^3_\R}'$.  We introduce two sets of scalars
\eqn\phiscalars{\phi^I = \phi^I_\CL(4,\bar{4},1) + \phi^I_\LR(1,4,\bar{4})
 + \phi^I_\RC(\bar{4},1,4)}
labeled by the flavor index $I=u,d$ to accomplish this breaking.  The
$\phi_\LR^u$ and $\phi_\LR^d$ fields are presumed to acquire the distinct
vacuum expectation values
\eqn\phiLRvevs{
\vev{\phi^u_\LR} = \pmatrix{0 & 0   & 0 & 0    \cr
		 	    0 & 0  & 0 & \vL  \cr
			    0 & 0   & 0 & 0    \cr
			    0 & \vR & 0 & \LLR \cr} \quad\qquad
\vev{\phi^d_\LR} = \pmatrix{0 & 0 &\vL   & 0 \cr
  		 	    0 & 0 & 0    & 0 \cr
			  \vR & 0 & \LLR & 0 \cr
			    0 & 0 & 0    & 0 \cr}.}
Heavy Dirac masses for the $U$, $N$, $D$ and $E$ fermions are
then generated via the Yukawa interaction
\eqna\Yukawatype
$$ \eqalignno{{\cal L}_{\rm Yukawa}(\phi^\I) =
f^I \Tr \bigl[ (\Psi^\T_\RC) (\bar{4},1,4) &C
 \Psi_\CL (4,\bar{4},1) \phi_\LR^I(1,4,\bar{4}) & \cr
+ (\Psi^\T_\CL) (4,\bar{4},1) &C
 \Psi_\LR (1,4,\bar{4}) \phi_\RC^I(\bar{4},1,4) & \Yukawatype a \cr
+ (\Psi^\T_\LR) (1,4,\bar{4}) &C
 \Psi_\RC (\bar{4},1,4) \phi_\CL^I(4,\bar{4},1) \bigr] + \hc & \cr} $$
We also give $O(\LLR)$ masses to all the new exotic leptons in
\familyIII{b}\ through a second Yukawa term
$$ \eqalignno{{\cal L}_{\rm Yukawa}(\chi)  =
{g \over 2} \Tr \bigl[ (\Psi^\T_\CL) (4,\bar{4},1) &C
 \Psi_\CL (4,\bar{4},1) \chi_\CL(6,6,1) & \cr
+ (\Psi^\T_\LR) (1,4,\bar{4}) &C
 \Psi_\LR (1,4,\bar{4}) \chi_\LR(1,6,6) & \Yukawatype b \cr
+ (\Psi^\T_\RC) (\bar{4},1,4) &C
 \Psi_\RC (\bar{4},1,4) \chi_\RC(6,1,6) \bigr] + \hc & \cr} $$
which antisymmetrically couples fermions to the additional Higgs
field
\eqn\chiscalar{X = \chi_\CL(6,6,1) + \chi_\LR(1,6,6) + \chi_\RC(6,1,6).}
The only components of $X$ that may develop nonvanishing vacuum expectation
values which do not break color and electromagnetism but do violate
$U(1)_{\L,\R}$ are $(\chi_\LR)^{[12]}_{[12]}$, $(\chi_\LR)^{[12]}_{[34]}$,
$(\chi_\LR)^{[34]}_{[12]}$ and $(\chi_\LR)^{[34]}_{[34]}$.  We choose these
VEV's to all equal $\LLR$.

	The final two symmetry breaking steps in \patternIII\ result from
the $\vR$ and $\vL$ entries in \phiLRvevs\ and
%
%
\eqn\phiRCLvevs{
\vev{\phi^I_\CL} = \pmatrix{0 & 0   & 0 & 0    \cr
		 	    0 & 0   & 0 & 0    \cr
			    0 & 0   & 0 & 0    \cr
			    0 & \vL & 0 & 0 \cr} \quad\qquad
\vev{\phi^I_\RC} = \pmatrix{0 & 0 & 0 & 0    \cr
		 	    0 & 0 & 0 & \vR  \cr
			    0 & 0 & 0 & 0    \cr
			    0 & 0 & 0 & 0 \cr}.}
The Yukawa Lagrangian induces mixing between the heavy seesaw fermions and
their light Standard Model partners.  The final forms of the quark and
charged lepton mass matrices appear as
\eqna\massmatrices{
$$ \eqalignno{M_{uU} &=
\bordermatrix{& u_\L & U_\L \cr
 u_\L^c & 0 & f^u \vR \cr
 U_\L^c & f^u \vL & f^u \LLR \cr} \qquad\qquad
M_{dD} =
\bordermatrix{& d_\L & D_\L \cr
 d_\L^c & 0 & f^d \vR \cr
 D_\L^c & f^d \vL & f^d \LLR \cr} & \massmatrices a \cr} $$
and
$$ \eqalignno{
M_{\rm \buildrel charged \over {\scriptstyle leptons}} &= \bordermatrix{
& e_\L & E_\L & I^-_\L & J^-_\L & K^-_\L & L^-_\L \cr
e^c_\L & 0 & f^d \vR & (f^u+f^d) \vL & 0 & 0 & 0 \cr
E^c_\L & f^d \vL & f^d \LLR & 0 & (f^u+f^d) \vL & 0 & 0 \cr
I^+_\L & (f^u+f^d) \vR & 0 & -g\LLR  & 0 & 0 & 0 \cr
J^+_\L & 0 & 0 & 0 & -g\LLR  & 0 & 0 \cr
K^+_\L & 0 & (f^u+f^d) \vR & 0 & 0 & -g \LLR & 0 \cr
L^+_\L & 0 & 0 & 0 & 0 & 0 & -g\LLR \cr}_. & \cr
&& \massmatrices b \cr} $$
We refrain from explicitly writing down the neutral lepton matrix since it
is larger and more complicated than those exhibited above.

	We should recall that the $f^u$, $f^d$ and $g$ Yukawa
couplings are $\NF \times \NF$ matrices in fermion family space.  As
we saw before in the $SU(5) \times SU(5)$ theory, it is useful to invoke a
parity symmetry $\bf{P}$ to constrain the forms of these Yukawa matrices.
We therefore follow ref.~\Carlson\ and promote the discrete $Z_3$ symmetry
in our $SU(4)^3$ model to $S_3$ through the addition of a parity
operation and its two cyclic partners.  $\bf{P}$ performs a
conventional spatial inversion and swaps the $SU(4)_\L$ and $SU(4)_\R$
factors in the gauge group.  Its action upon the $SU(4)_\C \times SU(4)_\L
\times SU(4)_\R$ gauge fields
\eqna\Paction
$$ \eqalignno{C^\u \xpt \to C_\u \mxpt \qquad
L^\u \xpt \to R_\u \mxpt \qquad
R^\u \xpt \to L_\u \mxpt && \Paction a \cr} $$
forbids a CP violating topological term from appearing in the gauge part of
the Lagrangian.  In the fermion sector, parity maps left handed fields into
their right handed analogs which we express as left handed conjugates:
$$ \eqalignno{\Psi_\LR \xpt &\to -C \bigl(\Psi_\LR^c \bigr)^* \mxpt & \cr
\Psi_\CL \xpt &\to -C \bigl(\Psi_\CL^c \bigr)^* \mxpt
 = -C\bigl(\Psi_\RC \bigr)^\dagger \mxpt & \Paction b \cr
\Psi_\RC \xpt &\to -C \bigl(\Psi_\RC^c \bigr)^* \mxpt
 = -C\bigl(\Psi_\CL \bigr)^\dagger \mxpt . & \cr} $$
{}Finally, the scalars transform under $\bf{P}$ as
$$ \eqalignno{
\eqalign{\Phi_\C \xpt &\to \Phi_\C^\dagger \mxpt \cr
\Phi_\L \xpt &\to \Phi_\R^\dagger \mxpt \cr
\Phi_\R \xpt &\to \Phi_\L^\dagger \mxpt \cr} \qquad
\eqalign{\chi_\LR \xpt &\to \chi_\LR^\dagger \mxpt \cr
\chi_\CL \xpt &\to \chi_\RC^\dagger \mxpt \cr
\chi_\RC \xpt &\to \chi_\CL^\dagger \mxpt \cr} \qquad
\eqalign{\phi_\LR^\I \xpt &\to \bigl(\phi_\LR^\I\bigr)^\dagger\mxpt \cr
\phi_\CL^\I \xpt &\to \bigl(\phi_\RC^\I \bigr)^\dagger \mxpt \cr
\phi_\RC^\I \xpt &\to \bigl(\phi_\CL^\I \bigr)^\dagger \mxpt. \cr} && \cr
&& \Paction c} $$
It is straightforward to check that the Yukawa interactions
in \Yukawatype{a,b}\ remain invariant under parity only if the $f^\I$
and $g$ coupling matrices are hermitian.  The fermion mass
matrices can thus be complex, but their determinants are real.  So
$\bar{\th} = \th_\QCD + \th_\QFD$ vanishes at tree level, and the
$SU(4)^3$ model provides a possible solution to the strong CP problem.

	We next diagonalize the fermion mass matrices in
\massmatrices{a,b}\ neglecting small intergenerational mixing between
families.  The masses of Standard Model quarks and charged leptons fix
the diagonal elements in $f^u$, $f^d$ and $g$:
\eqn\Yukcouplings{\eqalign{
f^d &\simeq {\LLR \over \vL\vR}
\pmatrix{m_d=0.009 && \cr & m_s=0.181 & \cr && m_b=4.5 \cr} \cr
f^u &\simeq {\LLR \over \vL\vR}
\pmatrix{m_u=0.005 && \cr & m_c=1.5 & \cr && m_t=130 \cr} \cr
g &\simeq {\LLR \over \vL\vR}
\pmatrix{  {(m_u+m_d)^2 \over m_e+m_d} && \cr
	 & {(m_c+m_s)^2 \over m_\u+m_s} & \cr
	&& {(m_t+m_b)^2 \over m_\tau+m_b} \cr} \cr}
\eqalign{
& = \sp \pmatrix{0.0007 && \cr & 0.0147 & \cr && 0.3654 \cr} \cr
& = \sp \pmatrix{0.0004 && \cr & 0.1218 & \cr && 10.556 \cr} \cr
& = \sp \pmatrix{ 0.0017 && \cr & 0.7998 & \cr && 233.9 \cr}. \cr}}
We have numerically evaluated these matrices using the indicated GeV quark
masses and the scale values in \scales.  The resulting Yukawa couplings for
the first and second families are reasonable in size.  We thus see the
seesaw mechanism at work generating small quark and lepton masses without
an excessive fine tuning of Yukawa couplings.  Unfortunately, the results
for the third family are corrupted by the huge top quark mass.  The
large value for $m_t$ can of course be offset by the inverted seesaw
prefactor in \Yukcouplings{}.  But then we are left with very
small Yukawas for the lightest quarks and leptons as in the Standard Model.

	{}Finally, we investigate proton decay in the $SU(4)^3$ theory.
Recall that left handed Standard Model fermions and antifermions
appear in separate multiplets in \familyIII{}.  Therefore, gauge
boson exchange cannot mediate fermion number violating transitions such
as $P \to \pi^0 e^+$.   Proton decay only proceeds through $\chi$ scalar
exchange graphs like the one illustrated in \pdecay.  We expect the mass
of the $\chi^\fourthirds$ scalar shown in the
figure to be on the order of the unification scale
$\Mgut= 6.47 \times 10^{11} \GeV$.  This mass seems
much too light to yield a proton lifetime consistent with the experimental
lower limit \PDB
\eqn\expPlifetime{\tau_\P > 5 \times 10^{32} \yrs .}
However, the diagram in \pdecay\ is further suppressed by $O(\vR/\LLR)^2$
as a result of seesaw mixing between fermion gauge and mass eigenstates.
Such seesaw suppresion of proton decay is generic in all unified seesaw
models.  Naive dimensional analysis yields the proton lifetime
estimate
\eqn\thPlifetime{\tau_\P \simeq {16\pi\over (g_{11})^4}
\Bigl( {\LLR \over \vR} \Bigr)^4 {M_\chi^4 \over m_\P^5}}
where $g_{11}$ is the Yukawa coupling for the first family in
\Yukawatype{b}\ while $16\pi$ represents a two body phase space factor.
Inserting numerical values, we find $\tau_\P \simeq 4.6 \times 10^{33} \yrs$
which is consistent with the bound in \expPlifetime.

\newsec{Conclusion}

	The $SU(5) \times SU(5)$, $SO(10) \times SO(10)$ and
$SU(4) \times SU(4) \times SU(4)$ models that we have investigated in this
paper illustrate the basic features of unified universal seesaw theories.
They also generalize several well-known models that have been studied in
the past.  Many possible extensions of this work would be interesting to
pursue.  Gauge boson mixing, neutrino masses and
loop contributions to $\bar{\th}$ should all be further analyzed in these
models.  Moreover, a number of alternatives to the symmetry breaking
patterns that we have considered here remain to be examined in the
$SO(10) \times SO(10)$ and $SU(4)^3$ theories.  Generalizations
to $E_6 \times E_6$ and $SU(5)^3$ which may maintain left-right symmetry
down to the Standard Model subgroup could also be constructed.  In short,
unified universal seesaw models represent a new class of Grand Unified
Theories in which there is much room for further exploration.

\bigskip
\centerline{\bf Acknowledgements}
\bigskip

	It is a pleasure to thank Howard Georgi and Sheldon Glashow for
numerous discussions in the past on many issues related to this
work.

\bigskip\bigskip\bigskip\bigskip\noindent

\listrefs
\listfigs
\bye